%% file: main.tex
\documentclass[conference]{IEEEtran}
\IEEEoverridecommandlockouts

\usepackage{cite}
\usepackage{amsmath,amssymb,amsfonts}
\usepackage{algorithmic}
\usepackage{graphicx}
\usepackage{textcomp}
\usepackage{xcolor}
\usepackage{url}
\usepackage{subcaption}

\def\BibTeX{{\rm B\kern-.05em{\sc i\kern-.025em b}\kern-.08em
    T\kern-.1667em\lower.7ex\hbox{E}\kern-.125emX}}
\begin{document}

\title{Power-Efficient RAN Intelligent Controllers Through Optimized KPI Monitoring}

\author{\IEEEauthorblockN{João Paulo S. H. Lima, George N. Katsaros, and Konstantinos Nikitopoulos}
\IEEEauthorblockA{\\University of Surrey, Guildford, United Kingdom \\ Emails: \{j.lima, g.katsaros, k.nikitopoulos\}@surrey.ac.uk}

}

\maketitle

\begin{abstract}
\input{abstract}
\end{abstract}


\section{Introduction}
\input{introduction}

\section{RIC Architecture Overview}
\input{arch-review}

\section{Experimental Setup}
\input{system-configuration}

\section{E2 Interface Traffic and Power Measurement}
\input{results-discussion}

\section{Addressing Redundant KPI Transmissions}
\input{kpi-selector}

\section{Conclusions}
\input{conclusions}

\section*{Acknowledgments}
This work has been supported by the “HiPer-RAN” project, 
winner of UK’s DSIT Open Networks Ecosystem Competition.
\bibliographystyle{IEEEtran}
\bibliography{IEEEabrv,refs}
\end{document}

%% file: abstract.tex
The Open Radio Access Network (RAN) paradigm envisions a more flexible, interoperable, and intelligent RAN ecosystem via new open interfaces and elements like the RAN Intelligent Controller (RIC). However, the impact of these elements on Open RAN's power consumption remains heavily unexplored.
This work for the first time evaluates the impact of Key Performance Indicator (KPI) monitoring on RIC's power consumption using real traffic and power measurements. By analyzing various RIC-RAN communication scenarios, we identify that RIC's power consumption can become a scalability bottleneck, particularly in large-scale deployments, even when RIC is limited to its core operational functionalities and without incorporating application-specific processes.
In this context, we also explore potential power savings through the elimination of redundant KPI transmissions for the first time, extending existing techniques for identical subscription removal and KPI selection. We achieve significant power consumption gains exceeding 87\% of the overall RIC power consumption.

%% file: introduction.tex
The Open Radio Access Network (RAN) introduces improved interoperability, network observability, and flexibility with new open interfaces and architectural components~\cite{polese2023understanding}.
Central to this framework is the RAN Intelligent Controller (RIC), which enables closed-loop automation and network optimization by supporting the deployment of applications tailored for various use cases (e.g., radio resource management, energy savings)~\cite{oran-survey-marinova}.
Despite the promise of Open RAN for a more intelligent and diversified RAN ecosystem, its impact on power consumption, particularly regarding the introduction of additional network elements (e.g., RIC), remains largely unknown.
The increased reliance on softwarization, along with the demands of real-time data analytics and automation, introduces uncertainties about how these factors will affect the system-level energy efficiency of Open RAN deployments~\cite{oran-energyawareness}.

The RIC plays a critical role in the operational efficiency and power footprint of Open RAN~\cite{deep-oran-review}.
Specifically, RIC enables the deployment of rApps and xApps, which are third-party applications designed to optimize various network functions, such as handover management, radio resource allocation, network security, and energy savings~\cite{oran-usecases-wg1}.
These applications depend on the periodic sharing of Key Performance Indicators (KPIs) from the network nodes (i.e., E2 nodes) managed by the RIC.
The applications use these KPIs to optimize or coordinate RAN operations.
To maximize its effectiveness in intelligent decision-making, RIC is expected to oversee multiple nodes simultaneously, with each node transmitting numerous KPIs.
In specific cases, these KPIs can be requested from each connected device to the network node, significantly increasing the volume of traffic that must be exchanged and processed~\cite{traffic-steering-lacava}.
As we discuss in this work, the resulting processing workload introduces significant computational demands, whose impact on power consumption can pose challenges not only to the RIC's energy efficiency but also to the overall scalability of Open RAN deployments.

Initial studies have primarily focused on the computational aspects of Open RAN~\cite{5gperf,towardgreener,deep-oran-review}, leaving the power implications of its additional architectural elements, particularly the RIC, largely unexplored.
Moreover, although~\cite{experimental-eval,datasets} provide power consumption measurements and datasets of Open RAN systems, these focus on virtualized RAN elements and do not cover RIC in the evaluations.
Furthermore, while it has been generally understood that the KPI traffic managed by the RIC impacts its computational workload, no prior research has quantified its effect on power consumption or assessed how specific KPI monitoring activities (e.g., duplicate subscription removal) can result in measurable power savings.

In this work, for the first time, we quantify the impact of KPI monitoring on RIC's power consumption through actual traffic and power measurements.
To achieve this, we generate diverse RIC processing workloads, by varying system parameters, such as the number of active E2 node subscriptions and the number of KPIs reported per node.
We find that both E2 traffic and the number of reported KPIs per node are linearly related to RIC's power consumption, highlighting significant power-efficiency challenges for larger-scale deployments. 
In this context, also for the first time, we evaluate the potential power savings achieved by eliminating redundant KPI transmissions, extending existing techniques such as duplicate subscription removal to a more finely-grained approach. Specifically, we propose an approach that selects and transmits KPIs collectively based on the reporting periods of all corresponding xApps. This proposed mechanism, as we show in this work, achieves RIC power savings that can exceed 87\% or 1.2~kW, also allowing for more scalable deployments.
Finally, focusing solely on the core functionalities required by the RIC, we establish lower-bound estimates of its power consumption. This provides a foundation for future extensions of RAN power models to account for RIC processing and enable further optimization tailored to specific xApp use cases.

The rest of the paper is organized as follows: Section~\ref{sec:arch-review} describes the RIC architecture, detailing the connections with other architectural elements and the main internal components.
Section~\ref{sec:methodology} presents the setup used for our evaluations.
In Section~\ref{sec:measurements}, we present the measurements of E2 interface traffic and power consumption for diverse RIC workloads. 
Next, we present in Section~\ref{sec:kpi-selector} the principles for KPI request optimization, including the results of the potential gains from redundant KPI transmission avoidance.
Finally, the conclusions and future works are presented within Section~\ref{sec:final-remarks}.

%% file: arch-review.tex
\label{sec:arch-review}

\begin{figure}
    \centering
    \includegraphics[width=0.99\linewidth]{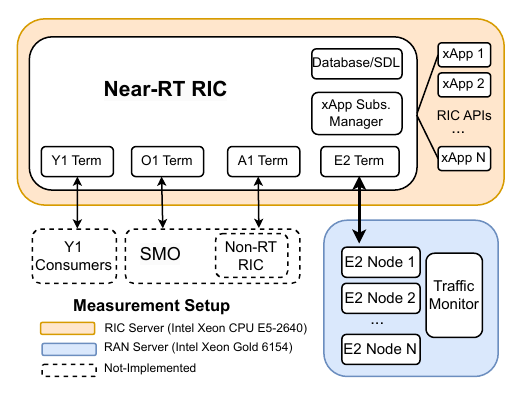}
    \caption{The Near-RT-RIC architecture, and the employed measurement setup, showing the key interfaces and interactions between the Near-RT RIC, Service Management and Orchestration (SMO), Non-RT RIC, Y1 consumers, and xApps.}
    \label{fig:oran-arch}
    \vspace{-10pt}
\end{figure}

This section discusses the RIC architecture, its main internal components, and the connections to other Open RAN elements, as seen in Fig.~\ref{fig:oran-arch}~\cite{oran-ricarch-wg3}.
The RIC platform is split into two components based on their corresponding latency requirements.
First, the Non-Real-Time RIC (Non-RT RIC) is designed for a control loop greater than 1 second, making it ideal for automating tasks such as network management, Artificial Intelligence (AI), and Machine Learning (ML) model training, and policy development.
Second, the Near-Real-Time RIC (Near-RT RIC) operates in the range of milliseconds, operating closer to the RAN and targeting optimization tasks through fast-paced analytics and inference~\cite{oran-survey-marinova,oran-ricarch-wg3}.

The internal components of Near-RT RIC 
include (1) its \textit{interface terminations}, allowing the integration with other elements; (2) the \textit{xApp Subscription Manager}, coordinating the initialization of xApps and the identification of potential identical subscriptions; (3) the \textit{Database and Shared Data Layer} (SDL), for managing the access to information acquired from RAN and produced by xApps; (4) the \textit{AI/ML Support} component, supplying data pipelining, model management and other tools for AI/ML operation~\cite{oran-ricarch-wg3}.

The interfaces for RIC integration are E2, A1, O1, and Y1, each targeting different parts of the architecture.
The E2 interface connects Near-RT RIC to the E2 nodes, which are RAN components such as the Centralized Units (CUs) and the Distributed Units (DUs).
Through E2 interface, the KPI exchange is established to feed RIC applications with RAN information.
Additionally, the RAN control and indication messages originating from those applications are redirected to the specific RAN nodes.
As such, E2 interface's traffic is key for the analysis conducted in this work.
The A1 interface establishes the communication between Non-RT and Near-RT RIC, aiming to share network policies and model management services.
Next, the O1 interface connects RIC and the Service Management and Orchestration (SMO) layer for specific software management and surveillance, as well as cloud resource scaling and administration.
Finally, the Y1 allows external consumers to access RAN information provided by the Near-RT RIC~\cite{polese2023understanding,oran-ricarch-wg3}.

The xApps, hosted on the RIC platform (Near-RT RIC), are software applications that utilize information gathered from the RAN to generate optimized outputs for specific use cases, employing advanced optimization algorithms such as AI/ML models. This enables automation and enhanced control over the RAN for use cases, including resource allocation, beamforming management, mobility coordination, energy efficiency, and security~\cite{oran-usecases-wg1}. The xApps interact with the RIC through Application Programming Interfaces (APIs), specifying the required KPIs and their reporting periodicity, as well as sharing the outputs generated by their operations.

%% file: system-configuration.tex
\label{sec:methodology}
In this section, we describe the experimental setup used for the evaluations.
As shown in Fig.~\ref{fig:oran-arch}, two servers connected via a 10 Gbps Ethernet link are utilized to run the RIC and RAN instances separately, ensuring accurate and reliable measurements.
The RAN server (Intel Xeon Gold 6154) runs E2 node instances from where the KPIs are shared to RIC, containing RAN information, in accordance with the O-RAN Alliance specifications.
These are the main traffic sources for our evaluations since the initial connection establishment represents only a small portion of the communication process, consisting of subscription request/response messages.
Upon the modification of different design parameters, such as the number of connected E2 nodes, it is possible to examine the performance of RIC in terms of E2 traffic, computational resources utilization, and power consumption, under diverse workloads. 
The RAN server is also responsible for monitoring the traffic from the E2 interface (downlink and uplink rates).
For that, we use the System Activity Reporter (SAR)~\cite{sar}, with a sampling interval of 1~second. 

\begin{figure*}[t]
\centering
 \begin{subfigure}[b]{0.49\columnwidth}
        \centering
        \includegraphics[height=6cm,width=0.99\columnwidth]{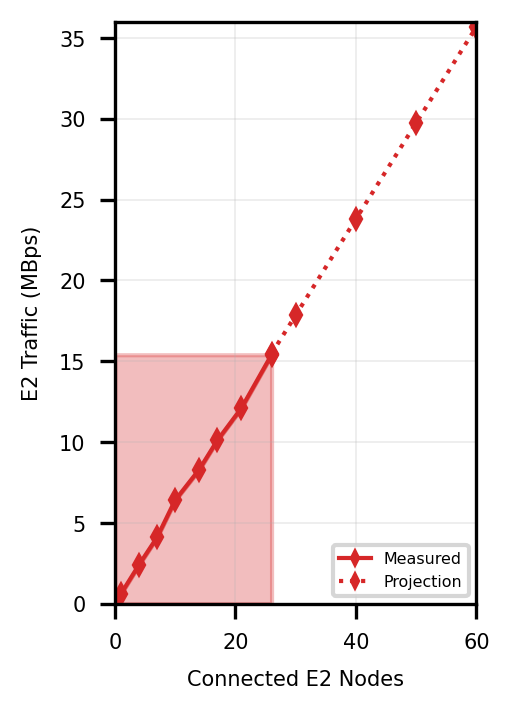}
        \caption{E2 traffic in Experiment 1}
        \label{fig:gnbs-traffic}
    \end{subfigure}
    \begin{subfigure}[b]{0.49\columnwidth}
        \centering
        \includegraphics[height=6cm,width=0.99\columnwidth]{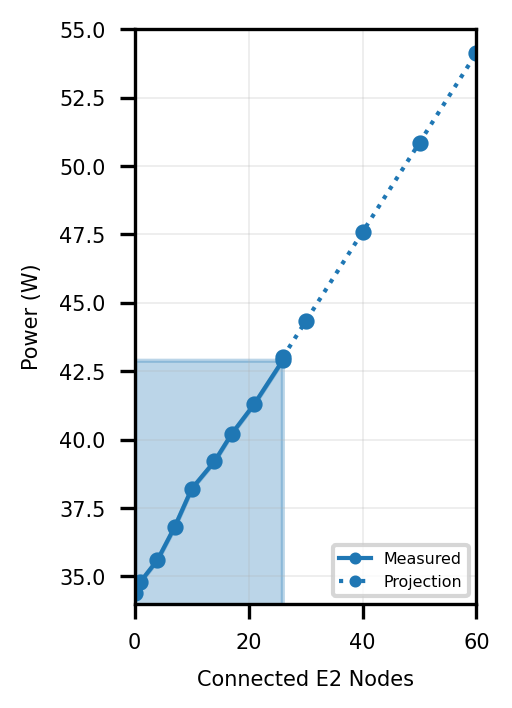}
        \caption{RIC power in Experiment 1}
        \label{fig:gnbs-power}
    \end{subfigure}
    \begin{subfigure}[b]{0.49\columnwidth}
        \centering
        \includegraphics[height=6cm,width=0.99\columnwidth]{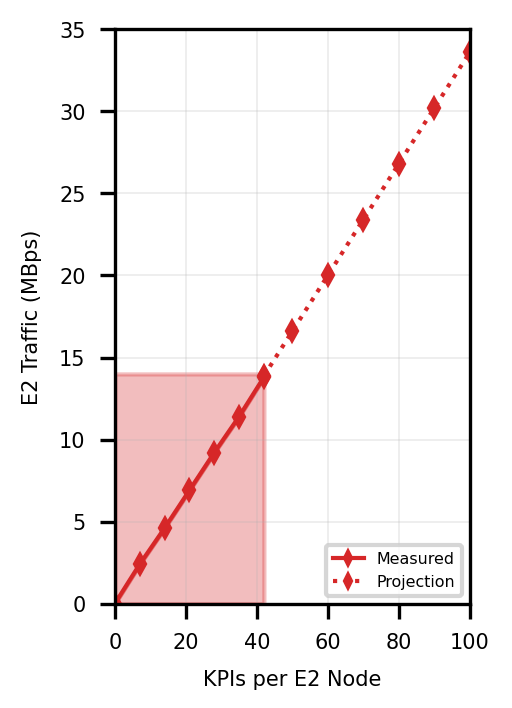}
        \caption{E2 traffic in Experiment 2}
        \label{fig:kpis-traffic}
    \end{subfigure}
    \begin{subfigure}[b]{0.49\columnwidth}
        \centering
        \includegraphics[height=6cm,width=0.99\columnwidth]{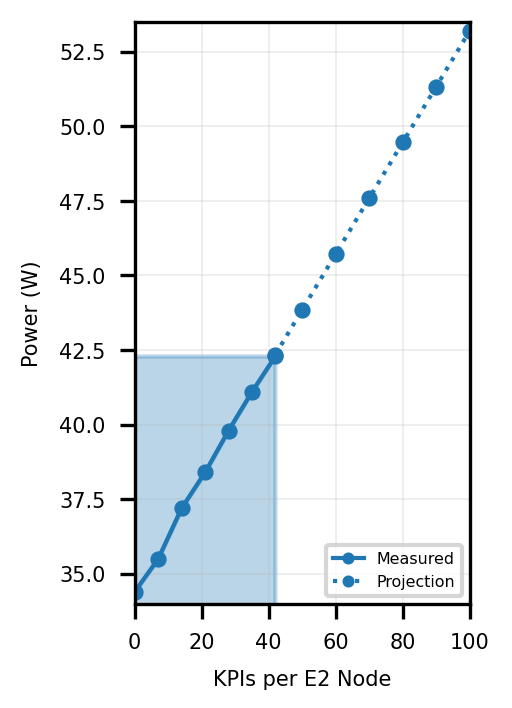}
        \caption{RIC power in Experiment 2}
        \label{fig:kpis-power}
    \end{subfigure}
    \caption{Evaluations of E2 traffic and RIC power for different workloads.}
     \label{fig:all}
     \vspace{-10pt}
\end{figure*}

The RIC server (Intel Xeon CPU E5-2640) hosts the RIC platform (Near-RT RIC) and the xApp instances.
We employ a modified version of OpenAirInterface (OAI) FlexRIC~\cite{flexric} that allows for a larger number of requests per second enabling the measurements of some computationally intensive experiments discussed in Section \ref{sec:measurements}.
The employed xApp is FlexRIC's KPI Monitoring application, which is designed as a simple requester and collector of RAN metrics, through messages compliant with O-RAN Alliance standards.
The RIC server also captures the CPU power consumption, following the measurement methodology of \cite{softiphy}, utilizing Turbostat, which samples
Intel's Running Average Power Limit Energy Reporting (RAPL) every second.

%% file: results-discussion.tex
\label{sec:measurements}

This section presents measurements of E2 traffic and power consumption for different RIC workloads, while also providing projections for larger deployments.
We perfrom two main evaluations, varying two parameters to produce the experimental workloads: the number of connected E2 nodes, and the number of KPIs shared per E2 node.
It is worth noting that for all evaluations the use of processors' C-states~\cite{cstates} in the RIC server is enabled via the server's BIOS configuration so that the effects of the experimented parameters are observable with power-saving states in action.

The first experiment was conducted by varying the number of connected E2 nodes while running one instance of the KPI Monitoring xApp sharing 7~KPIs with a constant report period of 10~ms (FlexRIC default configuration).
The employed xApp is an essential monitoring application, that requests and stores the received data from RAN nodes, through messages compliant with the O-RAN Alliance specifications. This allows us to focus only on the fundamental functionalities expected to be implemented in every deployment, essentially providing lower-bound estimates of the E2 traffic and the power consumption.

During the initial procedures to establish the E2 communication and the xApp initialization, there is an active exchange of packets between RAN and RIC, with messages such as \textit{E2 Setup Request/Response}, \textit{RIC Subscription Request/Response}~\cite{oran-ricarch-wg3}.
However, the effects of those on power are negligible since they represent a minor contribution towards the E2 traffic.
Fig.~\ref{fig:gnbs-traffic} shows the linear increment in traffic generated by the addition of E2 node connections.
When 26 E2 nodes are connected to the RIC, we measure over 15~MBps of traffic.
Finally, we project the traffic for larger deployments, in which up to 60 E2 nodes are connected.
As a result, the traffic can exceed 35~MBps in such conditions.

The resulting power from this experiment is seen in Fig.~\ref{fig:gnbs-power}.
The CPU static power consumption is 28~W in the RIC server, which corresponds to the minimum power that is always spent just by having the server on, without RIC-related operations. 
When RIC is running, the power without E2 traffic is near 34.5~W, we refer to this as the RIC static power consumption.
In addition, we observe a linear increment in power when increasing the number of connected E2 nodes.
From this, the projections indicate a consumption of nearly 55~W for 60 E2 connections, representing an increase of 60\%.

It is worth noting that the number of KPIs is susceptible to being modified by each application's needs.
This number can significantly rise, especially in scenarios where per-UE statistics are essential (e.g., mobility management)~\cite{traffic-steering-lacava}.
As such, in our second experiment, we measure the E2 traffic and the power consumption for different numbers of KPIs shared by each E2 node.
For this test, the report period is fixed at 10~ms and the number of E2 nodes is 4.
Fig.~\ref{fig:kpis-traffic} and~\ref{fig:kpis-power} demonstrate the behavior of the E2 traffic and power consumed in this test.
The projections, in this case, show the power exceeding 50~W (an increase of over 50\%) when having more than 80 KPIs shared per E2 node, while the traffic can reach up to 35~MBps.

Moreover, it is expected that the specific xApp realization and their employed optimization algorithms can pose an additional power overhead to the presented results. Assessing the impact of different algorithms and xApp realizations (whether or not relying on AI/ML) may have on power consumption remains beyond the scope of this work and conveyed to future research that involves AI energy efficiency strategies~\cite{green-ai}. 
In this context, the presented results demonstrate that
RIC, only with its basic functionalities (i.e., KPI monitoring), may represent a power bottleneck for large-scale Open RAN deployments, especially in scenarios involving a substantial volume of transmitted KPIs.

%% file: kpi-selector.tex
\label{sec:kpi-selector}

\begin{figure}[t]
    \centering
\includegraphics[width=\columnwidth,height=8cm]{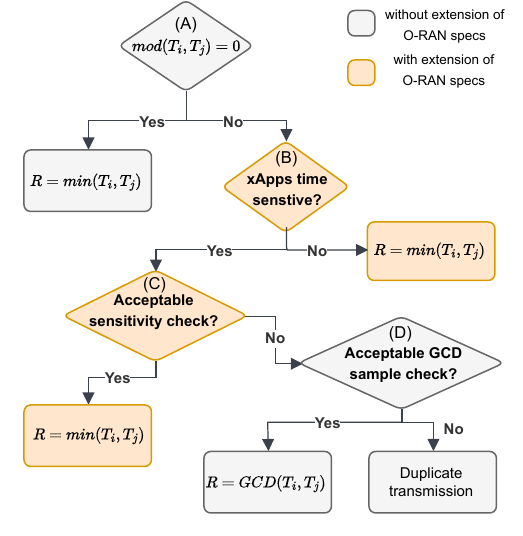}
    \caption{Principles of the KPI request merge process.}
    \vspace{-10pt}
    \label{fig:flowchart}
\end{figure}

This section presents the proposed mechanism for reducing redundant KPI transmissions, detailing its operational principles, integration within existing RIC architectures, and its impact on computational power consumption, including an evaluation of the potential gains.

\subsection{The Limitations of the Current Approaches}
\label{subsec:current-limitations}
The xApp Subscription Manager plays a central role in coordinating RAN data requests, as the centralizing entity for handling subscriptions.
This component analyzes the upcoming requests from xApps and merges identical E2 node subscriptions, reducing redundant E2 traffic~\cite{oran-ricarch-wg3}.
However, the mechanism for identifying the identical requests, realized by the O-RAN Software Community (OSC)~\cite{osc-page}, currently relies on comparing the MD5 hash functions of entire subscription requests.
This approach does not address scenarios where two requests partially overlap in a subset of the requested KPIs.
As demonstrated in Section~\ref{subsec:kpi-gains}, a more granular approach that accounts for such partial overlaps can yield significant reductions in both E2 traffic and RIC power consumption.
It is important to note that popular community solutions, such as OAI
FlexRIC, do not implement an O-RAN compliant Subscription Manager~\cite{oran-ricarch-wg3} to resolve such conflicts, even at the subscription request level. This may lead to multiple redundant copies of the same KPIs being transmitted, which in turn leads to proportionally increased power consumption, as shown in Section~\ref{sec:measurements}.

\subsection{Minimizing KPI Traffic}
\label{subsec:mechanism}
The proposed approach relies on breaking down upcoming subscription requests, so that each requested KPI is evaluated individually, and specific actions can be taken based on the current KPI monitoring activity. Specifically: 

\begin{itemize}
    \item If a KPI is requested to be shared with the same periodicity more than once, the mechanism should only ignore the new request, and provide access to the data that is already being shared.
    \item If a KPI is requested to be shared with a different periodicity than the one already in use, the selection should be made according to the following methodology, summarized in Fig.~\ref{fig:flowchart}.
\end{itemize}

The proposed mechanism aims to ensure that each requested KPI from a given E2 node is sampled and transmitted at the lowest sum rate possible, thereby eliminating redundant transmissions while meeting the requirements of all the involved xApps.
In cases where the different xApps request the same KPI on different report periods, this can be accomplished by simply sampling the corresponding KPI once, only at the smallest report period among all subscribed xApps requesting that KPI. 
For example, suppose xApp $i$ and xApp $j$ request KPI $a$ with report periods of $T_i= 10$ ms and $T_j=20$ ms, respectively. In that case, the proposed mechanism will merge the requests, ensuring that KPI $a$ is sampled and transmitted once, at the report period $T$ of 10~ms.

\begin{figure}[t]
    \centering
    \includegraphics[width=0.8\linewidth,height=8cm]{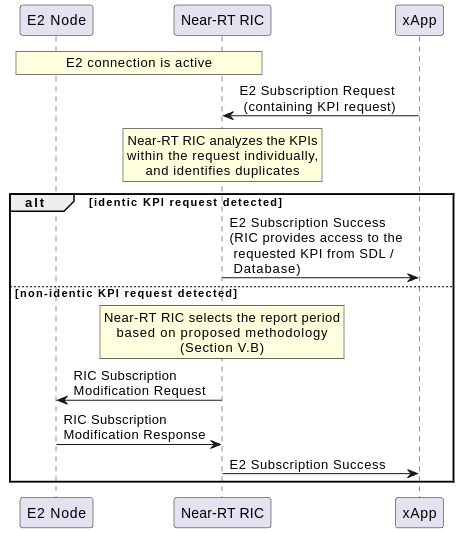}
    \caption{KPI selection mechanism detects potential KPI duplicates and optimizes the requests.}
    \label{fig:kpi-selection}
    \vspace{-10pt}
\end{figure}

\begin{figure*}[t]
  \centering
  \begin{minipage}[c]{0.33\textwidth}
    \begin{center}
      \subfloat[Small size: 10 E2 nodes, 20 KPIs]{\includegraphics[height=5cm,width=\linewidth]{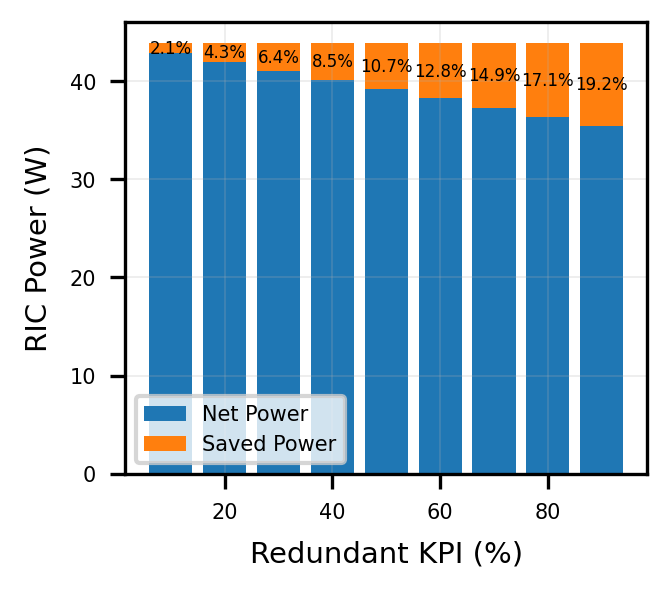}
        \label{fig:smallscale}}
    \end{center}
  \end{minipage}\hfill
  \begin{minipage}[c]{0.33\textwidth}
    \begin{center}
      \subfloat[Medium size: 100 E2 nodes, 50 KPIs]{\includegraphics[height=5cm,width=\linewidth]{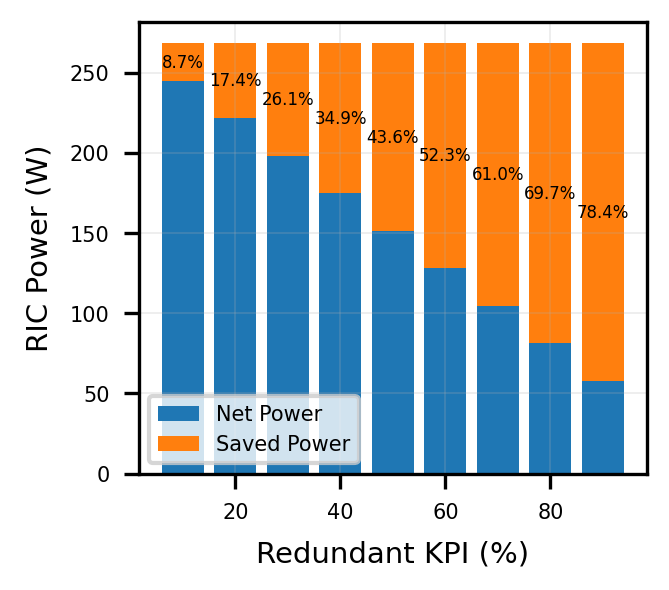}
      \label{fig:mediumscale}}
    \end{center}
  \end{minipage}\hfill
  \begin{minipage}[c]{0.33\textwidth}
    \begin{center}  
      \subfloat[Large size: 300 E2 nodes, 100 KPIs]{\includegraphics[height=5cm,width=\linewidth]{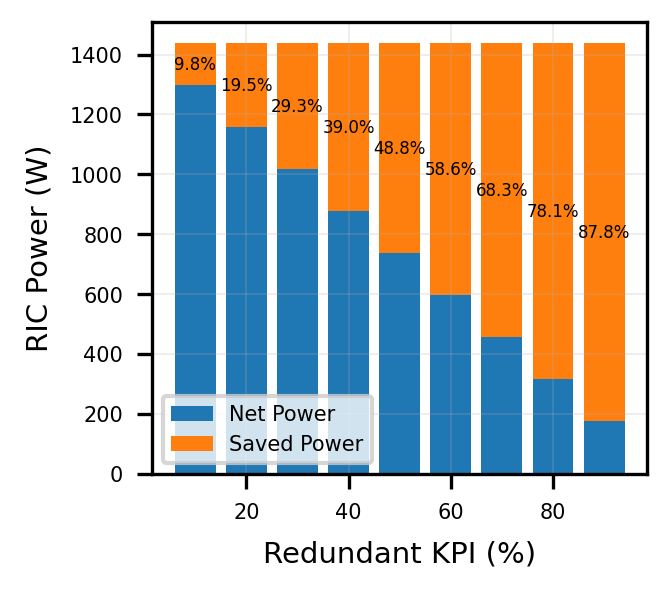}
      \label{fig:largescale}}
    \end{center}
  \end{minipage}
  \caption{The estimated RIC power savings with redundant KPI transmission avoided.}
  \label{fig:kpi-duplicate-savings}
  \vspace{-20pt}
\end{figure*}

However, challenges arise when the requested report periods are not divisible, i.e., $mod(T_i,T_j)\neq 0$ (A-Fig.\ref{fig:flowchart}), where $T_i,~T_j$ are the corresponding report periods of two xApps that request the same KPI from the same E2 node.
In such cases, sampling the corresponding KPI on the smallest report period may occasionally result in the xApp with the largest requested report period receiving outdated KPI samples.

For this reason, we introduce the concept of \textit{Temporal Sensitivity}, which dictates the allowance for a particular xApp to receive outdated samples of a specific KPI within a $\Delta t_m$ time interval from the requested report period.
For example, suppose xApp requests sampling of KPI $a$ with a report period $T$ and a temporal sensitivity of 1 second. In that case, it means that occasionally receiving KPI samples that are outdated by up to 1 second from the requested report period would not cause any operational issues.

In this context, the decision to allow the selection of the minimum report period (i.e., $T=min(T_i,T_j)$) for sampling a KPI relies on whether the potential time delay in receiving KPI samples from the ``fastest"-sampling xApp remains within the temporal sensitivity tolerance of the ``slowest"-sampling xApp for that specific KPI. This condition is satisfied when: $\Delta t_{max} < \Delta t_m$, 
(C-Fig.\ref{fig:flowchart}), where $\Delta t_m$ represents the temporal sensitivity threshold of the xApp whose sampling time is the largest, and $\Delta t_{max}$ is given by:
\begin{equation}
    \Delta t_{max} = min(T_i,T_j) - \mathrm{GCD}(T_i,T_j)
\end{equation}
where $\mathrm{GCD}(T_i,T_j)$ is the greatest common divisor of the corresponding report periods. This ensures that the misalignment of the KPI sampling does not introduce unacceptable delays relative to the xApp's temporal sensitivity.
It is worth noting here that the introduction of temporal sensitivity requires adjustments to the existing \textit{E2 Subscription Request} message structure to incorporate this additional parameter.

In the proposed approach, when temporal sensitivity information is not provided or the sensitivity test is unsuccessful, the acceptable solution would be either to merge the requests and transmit at the rate of \( \mathrm{GCD}(T_i, T_j) \) or to allow duplicate transmissions. The decision between these options depends on the number of samples required to be transmitted in each case (D-Fig.\ref{fig:flowchart}).
More specifically, we merge the requests into a single one that samples the KPI at \( T = \mathrm{GCD}(T_i, T_j) \) if the following condition is satisfied:
\begin{equation}
    S_{\mathrm{gcd}} < S_{T_i} + S_{T_j}
\end{equation}
where:
\[
S_{\mathrm{gcd}} = \frac{\mathrm{LCM}(T_i, T_j)}{\mathrm{GCD}(T_i, T_j)}
\]
is the number of samples required to be transmitted by the merged request over the period of the Least Common Multiple (LCM), and:
\[
S_{T_{i/j}} = \frac{\mathrm{LCM}(T_i, T_j)}{T_{i/j}}
\]
denotes the number of request messages required to sample the KPI at the \( T_i \) and \( T_j \) periods, respectively.

As a result, Fig.~\ref{fig:kpi-selection} demonstrates the improved coordination of the KPI requests, in the case of detecting KPI duplicates in the E2 communication. 
These principles can be employed without any major changes in Open RAN architecture, by encompassing the new parameter for Temporal Sensitivity into E2 Application Protocol (E2AP) procedures~\cite{oran-e2ap-wg3}, enabling the proposed KPI periodicity refinement in the xApp setup stages.
The remaining configurations (regarding the report period calculation itself) can be realized as an additional Subscription Manager functionality or a specific micro-service (e.g., an xApp).
Furthermore, any additional processing latency introduced by the proposed mechanism does not affect the latency constraints of the corresponding xApp operation, since this analysis takes place during the E2 subscription process, and thereby is carried out only once.

\subsection{The Power Gains of Optimizing KPI Transmissions}
\label{subsec:kpi-gains}

In this section, we assess the potential gains of the proposed efficient KPI exchange approach in RIC power consumption.
Specifically, we examine three indicative scenarios in which we vary the percentage of redundant KPI transmissions, each accounting for small, medium, and large-scale RAN deployments. 
As discussed in Section~\ref{subsec:current-limitations}, redundant KPI transmission can occur due to the Subscription Manager not identifying partially overlapping subscription requests.

In each scenario, we assess the gross and net power consumption, demonstrating the potential gains of the proposed approach.
The three examined scenarios are: a) a small-scale deployment of 10 connected E2 nodes sharing 20 KPIs each; b) an intermediary-scale deployment, with 100 connected E2 nodes sharing 50 KPIs each; and c) a large-scale deployment of 300 connected E2 nodes sharing 100 KPIs each.
The results are presented in Fig.~\ref{fig:kpi-duplicate-savings}.
In the first scenario, the RIC gross power is 43.8~W, with a dominant contribution from its static power (approximately 34.5~W, as shown in Section~\ref{sec:measurements}).
From that, as the percentage of redundant shared KPIs increases, the potential gains also increase and can nearly reach 20\%, representing a reduction of 8.4~W.
In the second scenario, the power and potential savings are significantly higher due to the more intense KPI monitoring activity.
The gross power is 268.2~W.
The potential savings, even with 10\% of redundant KPI transmission, is over 23~W (8.7\%) and can reach up to 210~W in worst cases, which is more than 6 times the RIC static power.
In our final evaluation, the RIC gross power exceeds 1.4~kW, and the large-scale scenario leads to substantial savings independently of the percentage of redundant KPI transmissions.
With only 10\%, more than 140~W can be saved, and the savings can scale up to over 1.2~kW (or 87\%).
According to~\cite{thesis-energy}, this is equivalent to the power of more than 80 RAN pico sites, or 15 micro sites.

Moreover, further reduction of redundant KPI transmissions can also be addressed with techniques that explore correlations and dimensionality reduction over the considered RAN data, such as the Principal Component Analysis (PCA)~\cite{dimension-reduction}.
With those techniques, a reduced number of inputs is needed to produce the same optimized output by the considered algorithms.
For instance, the works of~\cite{dimension-reduc-detectsys,boutiba-rlf} apply such techniques in security and radio link failure prediction applications for 5G networks.
It is worth mentioning that those techniques can work additively to the proposed E2 traffic reduction mechanism, jointly contributing to substantial energy efficiency improvements in future Open RAN deployments.

%% file: conclusions.tex
\label{sec:final-remarks}

This work evaluates for the first time the RIC operation through network traffic and power consumption measurements under diverse workloads.
By adjusting parameters such as the number of connected E2 nodes or the number of shared KPIs per E2 node, we produce different RIC workloads in an Open RAN testbed.
We demonstrate for the first time measurements for the traffic of the E2 interface 
and the RIC power consumption, observing a linear increase in power of up to 60\% with the developed experiments.
Moreover, we proposed a mechanism that enhances the E2 traffic and RIC power consumption by optimizing the KPI selection for RIC applications.
Specifically, redundant KPI transmissions are avoided through individual analysis of each KPI within subscription requests, merging identic transmissions when possible, or optimizing the report periodicity for non-identic requests.
The results demonstrate that the proposed solution can bring substantial gains in RIC power consumption of up to 1.2~kW or 87\% in the evaluated scenarios.
Future works encompass a holistic evaluation of power consumption in Open RAN systems, the energy efficiency of different AI/ML models when embedded in RIC applications, and potential strategies to further optimize RIC power consumption.